\documentclass[prd,showpacs,letterpaper,twocolumn,nofootinbib,longbibliography,floatfix,superscriptaddress]{revtex4-2}

\usepackage{CJKutf8}

\usepackage{tikz}
\usepackage{adjustbox}
\usepackage{amsmath}
\usepackage{graphicx}
\usepackage{fullpage}
\usepackage[colorlinks=true,allcolors=blue]{hyperref}
\usepackage[capitalise]{cleveref}
\usepackage[utf8]{inputenc}
\usepackage{siunitx}
\usepackage{array}

\DeclareSIUnit \s {\second}
\DeclareSIUnit \ns {\nano\second}
\DeclareSIUnit \mus {\micro\second}
\DeclareSIUnit \ms {\milli\second}
\DeclareSIUnit \MB {\mega\byte}
\DeclareSIUnit \GB {\giga\byte}
\DeclareSIUnit \TB {\tera\byte}
\DeclareSIUnit \PB {\peta\byte}
\DeclareSIUnit \Mbps {\mega\bit/\s}
\DeclareSIUnit \Gbps {\giga\bit/\s}
\DeclareSIUnit \Tbps {\tera\bit/\s}
\DeclareSIUnit \Pbps {\peta\bit/\s}
\DeclareSIUnit \kton {\kilo\tonne} 
\DeclareSIUnit \kt {\kilo\tonne}
\DeclareSIUnit \Mt {\mega\tonne}
\DeclareSIUnit \eV {\electronvolt}
\DeclareSIUnit \keV {\kilo\electronvolt}
\DeclareSIUnit \MeV {\mega\electronvolt}
\DeclareSIUnit \GeV {\giga\electronvolt}
\DeclareSIUnit \TeV {\tera\electronvolt}
\DeclareSIUnit \PeV {\peta\electronvolt}
\DeclareSIUnit \EeV {\exa\electronvolt}
\DeclareSIUnit \m {\meter}
\DeclareSIUnit \cm {\centi\meter}
\DeclareSIUnit \in {\inchcommand}
\DeclareSIUnit \km {\kilo\meter}
\DeclareSIUnit \kV {\kilo\volt}
\DeclareSIUnit \kW {\kilo\watt}
\DeclareSIUnit \MW {\mega\watt}
\DeclareSIUnit \MHz {\mega\hertz}
\DeclareSIUnit \mrad {\milli\radian}
\DeclareSIUnit \year {years}
\DeclareSIUnit \POT {POT}
\DeclareSIUnit \sig {$\sigma$}
\DeclareSIUnit\parsec{pc}
\DeclareSIUnit\lightyear{ly}
\DeclareSIUnit\foot{ft}
\DeclareSIUnit\ft{ft}
\DeclareSIUnit \ppb{ppb}
\DeclareSIUnit \ppt{ppt}
\DeclareSIUnit \samples{S}
\DeclareSIUnit \pe{PE}

\newcommand{\enu}{\E_\enu}

\usepackage{pifont}
\usepackage{enumitem}
\usepackage{url}
\usepackage{rotating}
\usepackage{epsfig,graphics,rotate}
\usepackage{flushend}
\usepackage{bm}
\usepackage{amssymb}
\usepackage{amsmath}
\usepackage{amsfonts}
\usepackage{gensymb}
\usepackage{booktabs}

\usepackage{microtype}
\usepackage{multirow, makecell}
\usepackage{lipsum}

\usepackage{xspace}
\usepackage{comment}
\usepackage{floatrow}
\usepackage{ragged2e}

\usepackage{nicefrac, xfrac}
\usepackage{mathtools} 
\usepackage{relsize}   

\usepackage{float}

\definecolor{lime}{HTML}{A6CE39}
\DeclareRobustCommand{\orcidicon}{
	\begin{tikzpicture}
	\draw[lime, fill=lime] (0,0) 
	circle [radius=0.16] 
	node[white] {{\fontfamily{qag}\selectfont \tiny ID}};
	\draw[white, fill=white] (-0.0665,0.095) 
	circle [radius=0.005];
	\end{tikzpicture}
	\hspace{-2mm}
}

\foreach \x in {A, ..., Z}{\expandafter\xdef\csname orcid\x\endcsname{\noexpand\href{https://orcid.org/\csname orcidauthor\x\endcsname}
			{\noexpand\orcidicon}}
   }

\begin{document}

\begin{CJK*}{UTF8}{gbsn}


\title{Two Watts is All You Need:\\ Enabling In-Detector Real-Time Machine Learning for Neutrino Telescopes \\ Via Edge Computing}

\author{Miaochen~Jin (靳淼辰)\orcidA{}}
\email{miaochenjin@g.harvard.edu}
\affiliation{Department of Physics \& Laboratory for Particle Physics and Cosmology, Harvard University, Cambridge, MA 02138, USA}

\author{Yushi~Hu (胡雨石)}
\affiliation{Department of Electrical and Computer Engineering \& Natural Language Processing Group, University of Washington, Seattle, WA 98195, USA}

\author{C.~A.~Arg{\"u}elles\orcidC{}}
\email{carguelles@fas.harvard.edu}
\affiliation{Department of Physics \& Laboratory for Particle Physics and Cosmology, Harvard University, Cambridge, MA 02138, USA}

\date{\today}

\begin{abstract}
The use of machine learning techniques has significantly increased the physics discovery potential of neutrino telescopes.
In the upcoming years, we are expecting upgrade of currently existing detectors and new telescopes with novel experimental hardware, yielding more statistics as well as more complicated data signals.
This calls out for an upgrade on the software side needed to handle this more complex data in a more efficient way.
Specifically, we seek low power and fast software methods to achieve real-time signal processing, where current machine learning methods are too expensive to be deployed in the resource-constrained regions where these experiments are located.
We present the first attempt at and a proof-of-concept for enabling machine learning methods to be deployed in-detector for water/ice neutrino telescopes via quantization and deployment on Google Edge Tensor Processing Units (TPUs).
We design a recursive neural network with a residual convolutional embedding, and adapt a quantization process to deploy the algorithm on a Google Edge TPU.
This algorithm can achieve similar reconstruction accuracy compared with traditional GPU-based machine learning solutions while requiring the same amount of power compared with CPU-based regression solutions, combining the high accuracy and low power advantages and enabling real-time in-detector machine learning in even the most power-restricted environments.
\end{abstract}
\maketitle
\end{CJK*}


\section{Introduction}\label{sec:Intro}
Neutrino telescopes are large-scale neutrino detectors built in naturally occurring media such as glaciers, mountains, lakes, and seas or even deployed in outer space. 
They aim to detect high-energy neutrinos produced in the collision of high-energy hadrons with ambient gas or radiation in astrophysical sources.
The steeply falling ($E^{-2.5}$~\cite{IceCube:2015gsk,IceCube:2020wum}) flux of these neutrinos, together with the smallness of the neutrino-nucleon cross-section~\cite{Formaggio:2012cpf}, has made the detection of these neutrinos challenging.
Approximately ten years ago, the IceCube Neutrino Observatory, a gigaton-ice-Cherenkov detector in Antarctica~\cite{IceCube:2016zyt}, discovered the diffuse emission of these neutrinos; see~\cite{Spiering:2012xe} for a historical review.

More recently, driven by reconstruction and event selection improvements made possible by the use of machine learning techniques, the IceCube collaboration has announced the detection of the first steady-state extragalactic neutrino source, NGC 1068~\cite{IceCube:2022der}, and the observation of our galaxy in neutrinos~\cite{IceCube:2023ame}.
These successes follow from prior searches for astrophysical neutrino sources, which, among other things, found evidence for emission from the TXS 0506+056 Blazar~\cite{IceCube:2018cha} in IceCube and hinted emission from our galaxy by ANTARES~\cite{ANTARES:2022izu}, a smaller neutrino telescope which was deployed in the Mediterranean sea.

Detecting and studying these neutrinos can provide unique information about the cosmic accelerators that produce them and potentially resolve the 100-year-old problem of the origin of cosmic rays.
Additionally, these neutrinos probe previously uncharted energy and distance regimes and thus constitute a unique probe of new physics see~\cite{Arguelles:2019rbn},e.g, Refs.~\cite{Bustamante:2018mzu,Arguelles:2015dca,IceCube:2021tdn,Shoemaker:2015qul,Bustamante:2016ciw,Song:2020nfh,Abdullahi:2020rge,Farzan:2018pnk,Reynoso:2022vrn,Arguelles:2023wvf,Arguelles:2019tum,Carloni:2022cqz,Arguelles:2022tki,Murase:2019xqi,Murase:2019tjj,Guepin:2022qpl} for specific examples.
Furthermore, the field of neutrino astrophysics is growing with two optical neutrino telescopes under construction: Baikal-GVD~\cite{Safronov:2020dtw} in Russia and KM3NeT~\cite{Adrian-Martinez:2016fdl} in the Mediterranean Sea. 
These are expected to be followed by next-generation optical neutrino telescopes such as IceCube-Gen2 in Antarctica~\cite{IceCube-Gen2:2020qha}, TRIDENT in China~\cite{Ye:2022vbk}, and P-ONE~\cite{Agostini:2020aar} in Canada; as well as a breath of Earth-skimming neutrino detectors which focus on finding evidence for tau neutrinos using Cherenkov light, TRINITY~\cite{Brown:2021lef}, particle showers, TAMBO~\cite{Thompson:2023pnl}, or radio, Grand~\cite{Alvarez-Muniz:2018bhp}.

These experiments are expected to produce a large amount of data: IceCube currently produces data at approximately 3~kHz with similar data rates expected at KM3NeT and Baikal GVD. In the meantime, high-accuracy algorithms process these data at a much lower rate, such that in order to achieve real-time processing, we need much more efficient algorithms. See Ref.~\cite{Yu:2023ehc} for a recent ML proposal to tackle these large rates.
The large data rate is expected to increase in next-generation experiments significantly, e.g., IceCube-Gen2 is expected to have eight times the data rate of IceCube, while TRIDENT will be thirty times larger.
For the detectors to tell interesting events apart from the backgrounds or to send real-time warnings to rare events, the need for real-time triggering and reconstruction algorithms becomes more prevalent.
Ref.~\cite{Yu:2023ehc} provides a solution based on sparse convolutions.
However, the problem of neutrino event reconstruction is not only that of large backgrounds; these algorithms also need to operate in resource-constraint environments.
For example, the IceCube detector Main Array operates Digital Optical Module (DOM) at 5.7 watts per module~\cite{IceCube:2016zyt}; additionally, other experiments, such as TAMBO or Grand, envision solar-power detection units that generate limited power. Under these restrictions, machine learning algorithms whose efficiency benefit from GPU parallelization cannot be deployed, and current real-time triggering algorithms are CPU-based fast regression that fits under the power restriction but are much less accurate: they serve only as a preliminary selection and more accurate reconstructions are performed off-line.

Edge computing refers to low-latency computing solutions that happen close to the source of the data, for example in real-time data processing situations. 
In 2018, Google announced an edge computing microarchitecture: the Edge Tensor Processing Unit (Edge TPU) Dev Board~\cite{TPU_Edge_intro}, which is a portable version of the TPU architecture that was developed and announced earlier~\cite{2017_tpu}. Inheriting the \textit{Matrix Multiplication Units} that enables fast machine learning inference from the TPU, this edge computing version runs inference on reduced size models, consuming only $3$ Watts of power in total for the Dev Board, with only $2$ Watts required by the TPU chip itself. 
The Edge TPU has since then enabled machine-learning-inference capability on many mobile computing devices and is under further development and optimization even today, see~\cite{chin2021highperformance} and~\cite{schaefer2023edge} for relevant discussions.
With a versatile architecture that allows for easy interfacing, compiling and deployments, and boosted with a software backend TensorFlow~\cite{tensorflow2015-whitepaper}, the Edge TPU becomes a suitable edge computing solution to achieving real-time in-detector machine learning inference for neutrino telescopes.

In this article, we introduce the first attempt at accelerating neutrino event reconstruction on edge computing devices using a recursive neural network (RNN) method with a residual convolutional input encoding, which enables an extremely low-power-consuming alternative to GPU-based machine learning algorithms. We will discuss the specific data pre-processing, network design and quantization procedure that makes possible the deployment of this algorithm on the Edge TPU.
In~\Cref{sec:Sim}, we introduce the detector geometries and data simulation used in this work; in~\Cref{sec:Hardware}, we briefly discuss the various hardware architectures we test our algorithm on and their reported power consumption specifications and layout, specifically the constraints of the Edge TPU hardware, serving as the motivation of discussions of the software methods; in~\Cref{sec:Software}, we lay out in detail the data pre-processing and network architecture in the context of edge-computing hardware limitations, and explain our methods and solutions, including a fine-tuning training procedure; in~\Cref{sec:Results}, we evaluate the accuracy and power performance of our approach; finally, in~\Cref{sec:Outlook}, we discuss the various future directions this work opens up, and encourage further exploration in the direction of low-power computing in neutrino detectors. 

\section{Detector and Data Simulation}\label{sec:Sim}
In this work, we will test the performance and accuracy of our network on two example detectors of the same geometry deployed in water and in ice, to which we assigned the names of WaterHex and IceHex respectively, where the hexagonal geometry is inspired by that of IceCube.
As with traditional and upcoming optical neutrino telescopes, the optical modules (OM) are arranged in vertical strings. 
The inter-string distance is set to 100~meters, and inter-OM distance along a string is set to 17~meters.
The arrays of each of the detectors constitute a total of 114 strings, which is similar to the expected KM3NeT final configuration~\cite{Adrian-Martinez:2016fdl}, with each string containing 60~OMs, summing to a total of 6840~OMs. 

\begin{figure}[h]
    \centering
    \includegraphics[width=0.9\textwidth]{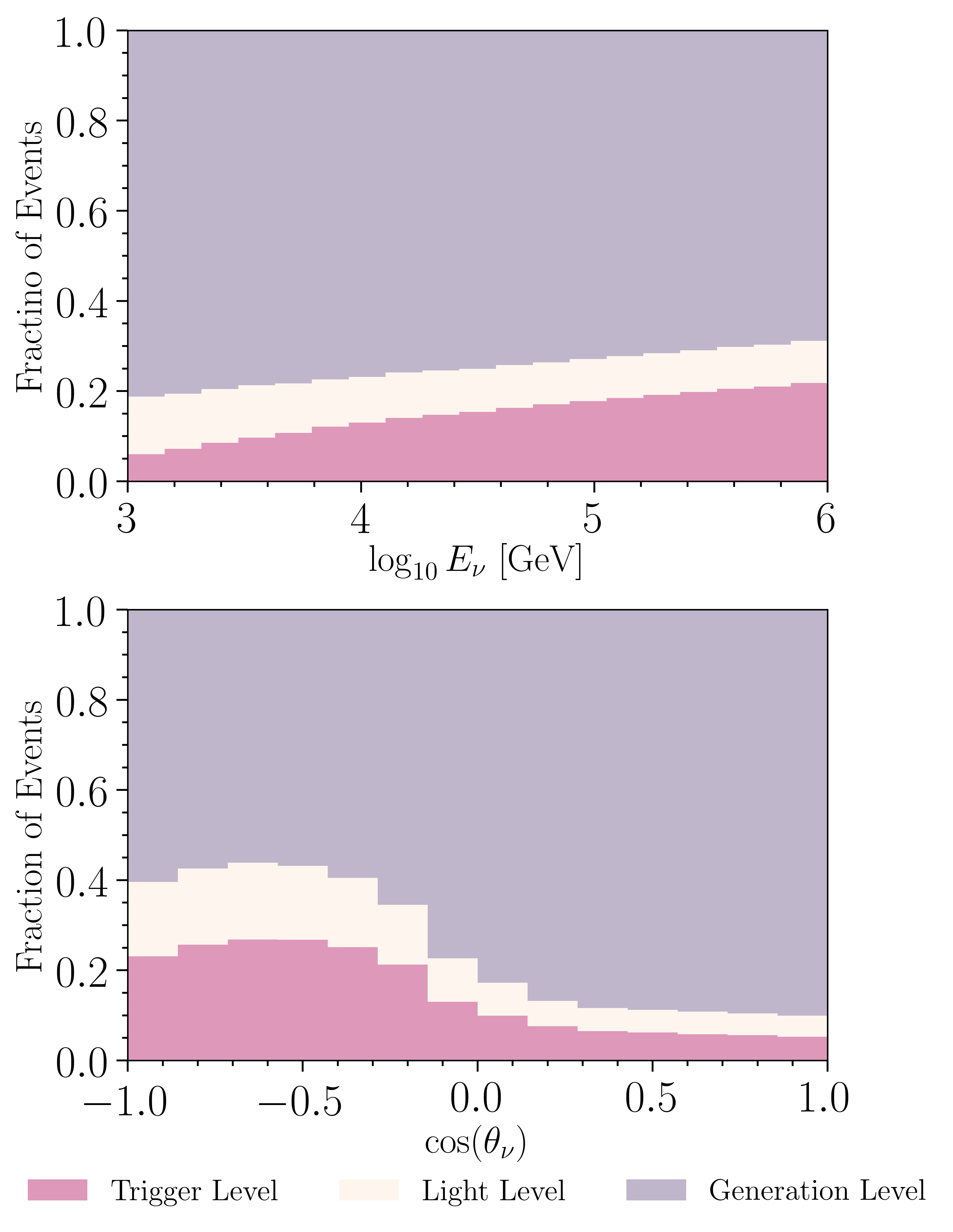}
    \caption{\textbf{Energy and Zenith angle distribution of data set at various selection levels} Plotted for simulated WaterHex dataset with the IceHex data set been qualitatively similar.}
    \label{fig:dataset}
\end{figure}

We use the open-source neutrino event simulation tool \texttt{Prometheus}~\cite{prometheus} to generate the neutrino events.
We generate all-sky neutrinos with $\cos \theta_{\mathrm{True}} \in [-1, 1]$ and $E_\nu \in [10^3, 10^6]$~GeV with a power-law energy distribution that has a spectral index of $-1$.
In~\Cref{fig:dataset}, we show the energy and cosine zenith distribution of the simulated events at different levels, the levels are to be understood as follows:
\begin{itemize}
    \item \textit{Generation level}: Includes all events generated by \texttt{Prometheus}.
    This is uniform in both energy and cosine of the zenith angle.
    In generating the events we use the volume injection option, where the neutrinos are injected such taht they pass through a uniform column depth sphere centered at the detector. More specifically, this spherical volume is defined such that column depth is uniform in all directions, with a cutoff at the top of the atmosphere. See~\cite{prometheus} for detailed documentation of this option setting.
    \item \textit{Light level}: Includes the events that contain at least one photon deposition in the array of OMs.
    The energy distribution as shown in~\Cref{fig:dataset} is due to the fact that larger energy neutrinos emit more photons and therefore are more likely to make photon deposits in a detector.
    The zenith distribution is skewed significantly towards the down-going direction, this is due to the aforementioned hard cutoff on event generation volume on the top of the atmosphere, leading to a much smaller column depth for down-going than for up-going muon tracks.
    This effect finally results in a significantly larger ratio of upgoing muon flux stopping before reaching the detector compared with its down-going counterpart, and thereby the skewed shape of the distribution.
    \item \textit{Trigger level}: This includes part of the light-level events that pass a trigger defined by a global simultaneous observation of local coincidences, similar to treatments preformed in other relavant works~\cite{Yu:2023ehc}.
    Specifically, we look for local coincidences observed within a time window of 5~ns by OMs that are on neighboring strings, and that are separated by at most 2~OMs apart in the vertical direction.
    Triggering is based on the detection of at least 8 local coincidences across the entire time window that the event spans.
\end{itemize}

Using these definitions, for down-going muon neutrinos, approximately $33\%$ deposit any light in the detector OMs, and about $25\%$ passes the trigger selection; for up-going muon neutrinos, on the other hand, only about $10\%$ deposits any light in the detector with $5\%$ passing the trigger selection.
At $1$ TeV, about $50\%$ of all simulated events deposit light in the detector OMs, and about $20\%$ of all events pass the trigger selection.
We simulated $2$ million events for the WaterHex detector and $2.5$ million events for the IceHex detector, out of which $300787$ and $373079$ events passed the trigger selection, respectively, both reaching about $15\%$ passing rate.
To facilitate cross-comparisons, we randomly choose $300000$ events for each detector and split the set of events randomly into $240000$ and $60000$ as their respective training and validation datasets.


\section{Hardware Setup}\label{sec:Hardware}
\subsection{Overview of Architectures and Power Consumption}

\begin{table*}[t]
\begin{tabular}{l|c|c|c}
\makecell{Hardware Architecture} & \makecell{Reported Efficiency} & \makecell{Total Power} & \makecell{ML Accelerator Power} \\
\hline\hline
\makecell{A100 GPU on Lenovo Server }	& 165 TFLOPS	&	2400W  &   400W\\
\hline
\makecell{RTX3080 GPU on Alienware Workstation }	& 29.8 TFLOPS	&	1000W  &   320W\\
\hline
\makecell{Apple MacBook Pro with M1 Pro chip}	& 5.3 TFLOPS	&	100W  &   15W\\
\hline
\makecell{Google Edge TPU }	& 4TOPS	&	3W  &   2W\\

\end{tabular}
\caption{\textbf{Summary of the hardware utility data used in the evaluation of the model deployed by this work.}
The power consumption specifications listed above are approximate values of peak consumption and therefore should be taken as a rough estimate and indication of the scale of the computing system to some extent of accuracy.
Even considering this limitation in precision, the comparison listed above still shows significant differences between different classes of hardware architectures.}
\label{table:hardware}
\end{table*}

For this work, we evaluate the performance, both accuracy and power efficiency, of our method on four different classes of hardware architectures spanning three different orders of magnitude in maximum power consumption, each serving as the representative of a class of computational architecture, including server/cloud scale, workstation/desktop scale, laptop scale, and edge computing scale.
In~\Cref{table:hardware}, we summarize the relevant specification parameters of the different hardware architectures: specifically, we show the reported Trillion Operations per Second (TOPS) / Trillion Floating Point Operations per Second (TFLOPS), total power, and ML accelerator power consumption. ``ML accelerator power'' counts only that of the GPU or TPU chip, whereas ``total power'' includes the consumption of the accompanying CPU and other parts of the cluster, PC or Edge TPU Dev Board respectively for the hardware architectures.

\subsection{The Google Edge TPU}

Specifically of interest to this article is the inference performance on the mobile, low-power machine learning accelerator: the Google Edge TPU. We devote this subsection to discussing the capabilities and limitations of the Edge TPU, which, as shown in~\Cref{table:hardware}, consumes only 2 Watts of power, as contrasted to large computing center-scale GPU clusters. 

The TPU architecture is capable of such efficient performance thanks to Matrix Multiplication Units (MXU), which are systolic arrays, in place of Arithmetic Logic Units (ALU), which are employed by CPU and GPU architectures. On the one hand, this architecture, while applied on server-level scale, is capable of running at 68$\times$ incremental performance per watt compared to GPU-based servers~\cite{2017_tpu}.
On the other hand, the Edge TPU architecture runs on integer operations instead of floating point operations and is optimized for low-power, mobile computing, an important requirement for enabling in online reconstruction work for neutrino telescopes and other experiments alike, often found in power-limited environments. While the Edge TPU inherits the performance advantages of its server-scale TPU relative, the capabilities come with trade-offs and harsh limitations on the software that can be run on it. While the Edge TPU documentation has an extensive list of requirements, including enabled layer and operation types~\cite{TPU_Op}, the two main restrictions that are crucial to the work of interest are as follows:

\begin{itemize}
    \item \textit{Dimensionality}: All tensors are restricted to $D \leq 3$, and in case where $D > 3$, the extra dimensions can only have length $1$. This also implies convolutional layers are restricted to $D \leq 2$, which is very sup-optimal for optical module array type detectors where the data is inherently 4 dimensional counting array and the time axis. Furthermore, the inability of the Edge TPU to handle high dimensional inputs disable batching, thereby resulting in expensive operations if one attempts to reduce 3D convolution to a batch of 2D convolutions. See~\Cref{subsec:LSTM} for a detail discussion.
    \item \textit{Quantization}: For full utilization of the Edge TPU requires \texttt{uint8} type accuracy instead of floating point accuracy. This means mapping all input, output, as well as network weights to integers within the range of $0$ to $255$. As one would expect, this means decreased accuracy for the model, but luckily, benchmark examples show that with proper treatments, quantized networks can also achieve very high accuracy performances~\cite{2022tpu_eval}. Various studies have also explore quantization methods that minimally reduce accuracy for different algorithms~\cite{chin2021highperformance}~\cite{chen2021quantization}~\cite{jacob2017quantization}
\end{itemize}

To overcome the hardships brought about by the limitations, we adapt our architecture design and employ a quantization procedure: these will be explored in~\Cref{subsec:LSTM} and~\Cref{subsec:quantization} respectively.

\section{Software Methods}\label{sec:Software}
\subsection{Data and Network Input} \label{subsec:data_input}

\begin{figure}[h]
    \centering
    \includegraphics[width=0.9\linewidth]{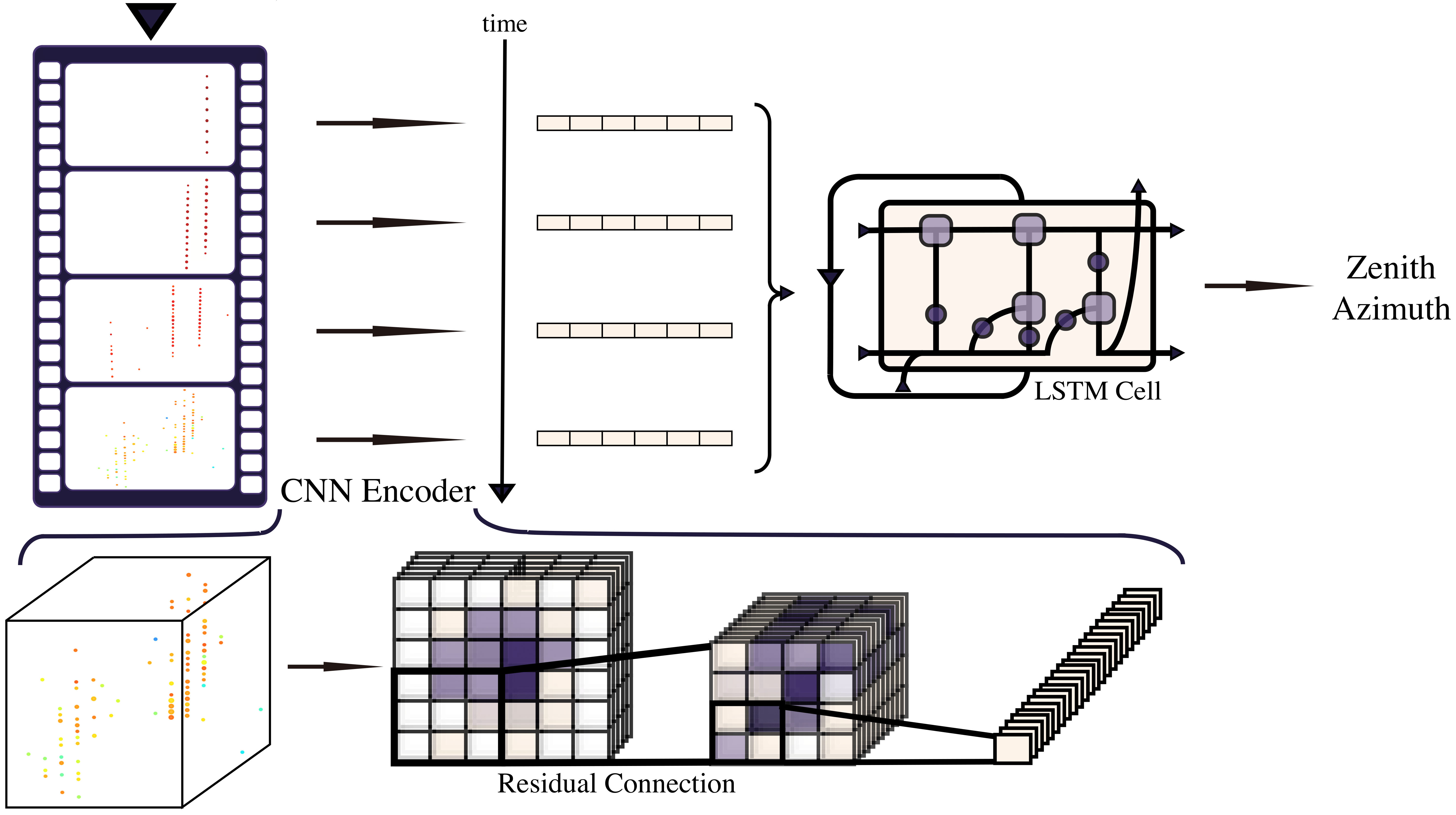}
    \caption{\textbf{Recursive Neural Network with Convolutional Encoding:} After input pre-processing, each time-step of the input data is encoded by a CNN encoder before processed by the LSTM cell to generate final prediction of the neutrino zenith and azimuth angles.}
    \label{fig:archi}
\end{figure}

Water(Ice)-Cherenkov detector data consists of the lowest level PMT waveforms recorded by optical modules in water or ice.
This yields a set of waveforms ($\{q_\alpha(t)\}_\alpha$) corresponding to each OM located at $(x_\alpha, y_\alpha, z_\alpha)$.
For machine learning algorithms to work with these data, at least some extent of pre-processing is needed, the specific method of which differs per ML algorithm and architecture employed.
We take IceCube as an example, where the prominent reconstruction algorithm \texttt{DNNreco}~\cite{IceCubeDNN} is a Convolutional Neural Network (CNN) approach, where data is pre-processed to form a 4-dimensional tensor, treated equivalent to an ``image.''
Spatial coordinates are embedded into a 3-dimensional tensor $(x_\alpha, y_\alpha, z_\alpha) \rightarrow V^{(i_\alpha, j_\alpha, k_\alpha)}$, while for the time component, the PMT waveforms $q_\alpha(t)$ are extracted to give nine distinct temporal features from the waveform of a DOM across the time of an event that describe the shape of the waveform. See~\cite{IceCubeDNN} for a detailed discussion. 
This results in an input of the format $\vec{T_i} \oplus \vec{x_i}$. With such an approach, the network primarily sees the data as an image, where on each ``pixel," now an OM location, the information of ``color channels" is replaced by the discrete time parameters.
In this work, we rethink the formulation of this problem, phrasing it primarily as a time series analysis and use Recursive Neural Network (RNN) architectures; this architecture choice will be elaborated upon in~\Cref{subsec:LSTM}.
We take the detector data after pulse extraction to obtain the individual hit times of photons on OMs, $\{H_i = (t_i, x_i, y_i, z_i)\}_{i=1}^N$, and group them imagining that we are taking snapshots of the event at a fixed timestep.
For an event consisting of $N$ hits that spans $\Sigma$ nanoseconds, we break it into $T$ separate frames each with $\sigma = \Sigma/T$ nanoseconds, containing an aggregate of $\{N_t\}_{t = 1}^{T}$ hits with $\sum_t N_t = N$.
This results in $T$ distinct three-dimensional arrays $A_t$, with each entry $A_t^{(X, Y, Z)}$ encoding the number of total hits on the corresponding DOM within the time window. Here $X, Y, Z$ are the three dimension of the OM array, wuch that the total number of photon hits in the entire detector within any specific time window is $\sum_{(x, y, z) = (1, 1, 1)}^{(X, Y, Z)} A_t^{(x, y, z)} = N_t$.
This data pre-processing results in a network input that resembles "snapshots" of the detector at different times, as shown in~\Cref{fig:archi}.

Another advantage of such a data representation is the discrete nature of all the entries in $A_T^{(X, Y, Z)}$.
For the network to be deployed on the TPU, not only the weights but also the network inputs have to be cast as full integers in the range of $[0, 255]$.
This input encoding design allows us to adapt the input to type \texttt{uint8} without dealing with errors that originate from mapping a continuous distribution to a discrete one to suit the Edge TPU model requirements.
A more detailed discussion on the full quantization of the network input will follow in~\Cref{subsec:quantization}.

\subsection{Recursive Network with Convolutional Embedding} \label{subsec:LSTM}

In most currently deployed machine learning-based reconstruction algorithms for neutrino telescopes, CNNs and graph neural networks (GNNs) are employed.
Given the data encoding, these architectures are straightforward and intuitive and yield satisfactory results. However, under the restrictions discussed in~\Cref{sec:Hardware}, especially the restriction on input tensor dimension and allowed types of operations, these architectures cannot be deployed on the Edge TPU device.

In~\cref{fig:archi} we show the network architecture developed in this work.
This is a combination of residual convolutional and recursive neural network architectures.
The design can be effectively summarized as an LSTM time-series prediction using residual convolution as the input encoder.
For each input array at timestep $t$, the CNN encoder transforms the array of image-like data $\{A_t^{(X, Y, Z)}\}_{t = 1}^T$ into a sequence of vectors $\{L_t\}$ in the latent space, which becomes the input into the hidden layers of the RNN.
The CNN encoder contains one initial convolution layer with five residual convolution blocks, each containing two convolution layers.
The LSTM cell contains an LSTM layer and a dense layer that connects to the output. The output is the azimuthal and zenith angles of the incoming neutrino, from which we calculated the error by a dot product with the simulated true neutrino direction.
See~\Cref{appx:NetworkArchi} for the detailed layout of the network architecture.

It is worth noting that in the CNN encoder, we employed a very peculiar way of treatment that takes apart a 3-dimensional array of OM photon hits $A^{(X, Y, Z)}$ and breaks it apart into $\{A^{(X,Y)}\}_{z = 1}^Z$, where the $z$ component is treated as channels in the network.
This is due to TPU's restriction on vector dimensions and time complexity considerations.
On the one hand, it is impossible to apply 3-dimensional convolution to the input data.
On the other hand, time complexity without parallelization capability forbids us to apply CNN on the $z$ component in parallel: an alternative way to the channel treatment is applying the same two-dimensional CNN on all vertical slices $\{A_z^{(X,Y)}\}$ to obtain a set of output vectors and concatenate them, followed by another 2D CNN that transforms the resulting 2D array into a 1D vector, wrapping up the encoding process.
While for a GPU, parallelization can be applied to the first set of CNN, and the entire process will consume the same time required for only two separate CNN networks, on the TPU architecture, where such parallelization is not allowed, this will take the same time as evaluating $(T+1)$ CNN networks separately.
Due to these constraints and considerations, we resorted to the design shown in~\Cref{fig:archi}.
We have tested the channel treatment versus the parallel treatment and found negligible differences in angular reconstruction accuracy on GPU architectures. 

\subsection{Quantization Procedure} \label{subsec:quantization}

Quantization for Edge TPU deployment is a necessary step where all network operations are cast into 8-bit full integer operations.
This is a daunting process where we face a trade-off between efficiency and accuracy.
We use TensorFlow's mobile library to realize the quantization process. 

To optimize for unsigned integer type operations, an important step during the conversion from the full precision model to the reduced precision one, the Edge-TPU compatible model must provide the converter with a representative dataset.
This is a set of input data, which we take from the training set, that helps the converter decide a reasonable mapping between $\texttt{float32}$ and $\texttt{uint8}$ by providing information regarding the range and distribution of the input, weights, and output.
Therefore, it is important to have a set of inputs with similar distributions and a narrow spread in value to ensure a smooth conversion to the reduced precision format.

We initiate the quantization process by focusing on the inputs. Referring back to our earlier discussions in~\cref{subsec:data_input}, the detector data is encoded using only integers. 
Notably, there are key features essential to the quantization process.
Firstly, as each entry represents the number of photons deposited within a sufficiently small time frame, only a minute fraction of entries surpass a certain threshold $A_t^{(x, y, z)} \leq 255$. 
Consequently, implementing a cutoff at 255 does not significantly impact the network's performance.
Secondly, in addition to a simple cutoff, we uniformly map the entries of every input tensor to a Gaussian distribution.
This approach proves beneficial for the quantization process as it ensures a more evenly distributed dataset. 

The input quantization process is summarized as
\begin{equation}
    x' = \mathrm{Clip}_{[0,255]}\Bigl(\texttt{uint8}(F_{(\mu, \sigma)}(x))\Bigl),
\end{equation}
where $F_{(\mu, \sigma)}$ maps the entire population $\{A_t^{(x, y, s)}\}_{(T, X, Y, Z)}$ to a Gaussian distribution centered at $\mu = 90$ with standard deviation $\sigma = 45$.
The choices of $\sigma$ and $\mu$ are selected after several trials such that the data is widely spread enough but minimally exceeding the range $[0, 255]$.
Although this pre-process quantization of the data is observed to impact the accuracy performance of the network negatively, the original network that's used for quantization and deployment on the TPU will be trained directly on this processed dataset.
In contrast, the same network for GPU-based training and inference will be trained on the original data input; a comparison between the two will be provided in~\cref{subsec:accuracy}.

Ideally, quantization of the network weights will be performed by the converter from the TFLite library; however, stable releases of TensorFlow do not yet support full quantization of networks with multiple subgraphs, such as the network developed by this work.
Instead, we divide the quantization process into steps to accomplish a successful conversion.
In each stage, we cast additional weights, activations, other layer components and outputs to an 8-bit unsigned integer.
These stages proceed as follows:
\begin{itemize}[label=\ding{254}]
    \item \textit{\textbf{Stage 1:} Quantization of Inputs.} To begin with, we have the inputs into the CNN encoder as $\texttt{uint8}$; all weights are $\texttt{float32}$.
    \item \textit{\textbf{Stage 2:} Quantization of CNN Encoder.} We quantize the CNN encoder by providing a representative dataset that contains 1000 input samples.
    However, after the CNN encoder evaluation, we use the reverse mapping to ``fallback'' from $\texttt{uint8}$ results to $\texttt{float32}$ numbers, which then becomes the input to the LSTM.
    Worth noting is that at this stage, while the input into the LSTM is indeed $\texttt{float32}$, there exists only $256$ distinct numbers, as they were mapped back from $\texttt{uint8}$-quantized CNN encoder outputs.
    At this stage, since the LSTM is unaware of the change of precision of the CNN Encoder, even in the best-case quantization scenario, we would not be able to recover the original reconstruction accuracy fully.
    This step, which is unnecessary if direct quantization of a multiple subgraph model is allowed, brings about a loss of accuracy that is avoidable with future software developments.
    \item \textit{\textbf{Stage 2.5:} Re-train the LSTM cell.} At this stage, we perform a fine-tuning re-training of the LSTM cell where the initial weights are directly transferred from the pre-trained original model as a solution to the artificial accuracy loss problem discussed in the previous bullet point.
    However, instead of training with the float-fallback CNN encoder outputs, we directly perform training on the quantized output from the CNN encoder, which is of type $\texttt{uint8}$.
    Upon finishing the re-training, the CNN encoder is fully quantized, and so is the input to the LSTM cell, but the LSTM cell still contains $\texttt{float32}$ weights.
    \item \textit{\textbf{Stage 3:} Quantization of LSTM and Dense Layers.} At this final stage, we perform the quantization of the LSTM cell, providing a representative dataset that contains 1000 samples of CNN encoder outputs of inputs selected from the training sample.
    The output is naturally also represented by the network in $\texttt{uint8}$, which falls back to $\texttt{float32}$ to provide us with the actual final prediction.
\end{itemize}
In~\Cref{sec:Results}, we will show the network performance at each stage.


\section{Results and Discussions}\label{sec:Results}
\subsection{Reconstruction Accuracy} \label{subsec:accuracy}

\begin{figure}[h]
    \centering
    \includegraphics[width=0.85\linewidth]{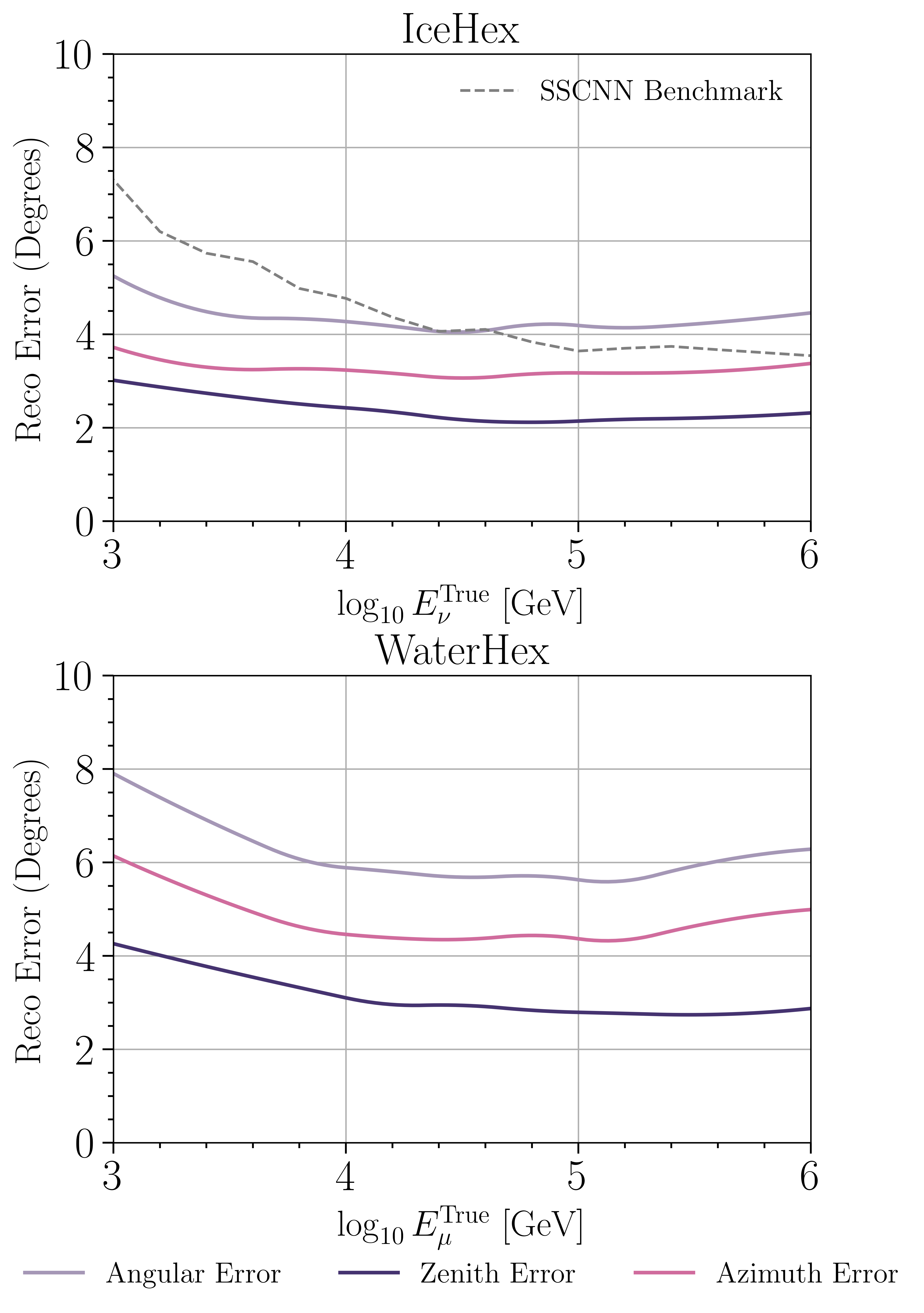}
    \caption{\textbf{Neutrino Angular Reconstruction Resolution on full accuracy network as a function of the true simulated neutrino energy:} Top panel shows the performance of the network trained on the IceHex detector, bottom panel shows the performance of the network trained on the WaterHex Detector. The SSCNN angular resolution is shown on the IceHex performance plot as a benchmark.}
    \label{fig:angular_reco_errors}
\end{figure}

\begin{figure}[h]
    \centering
    \includegraphics[width=0.85\linewidth]{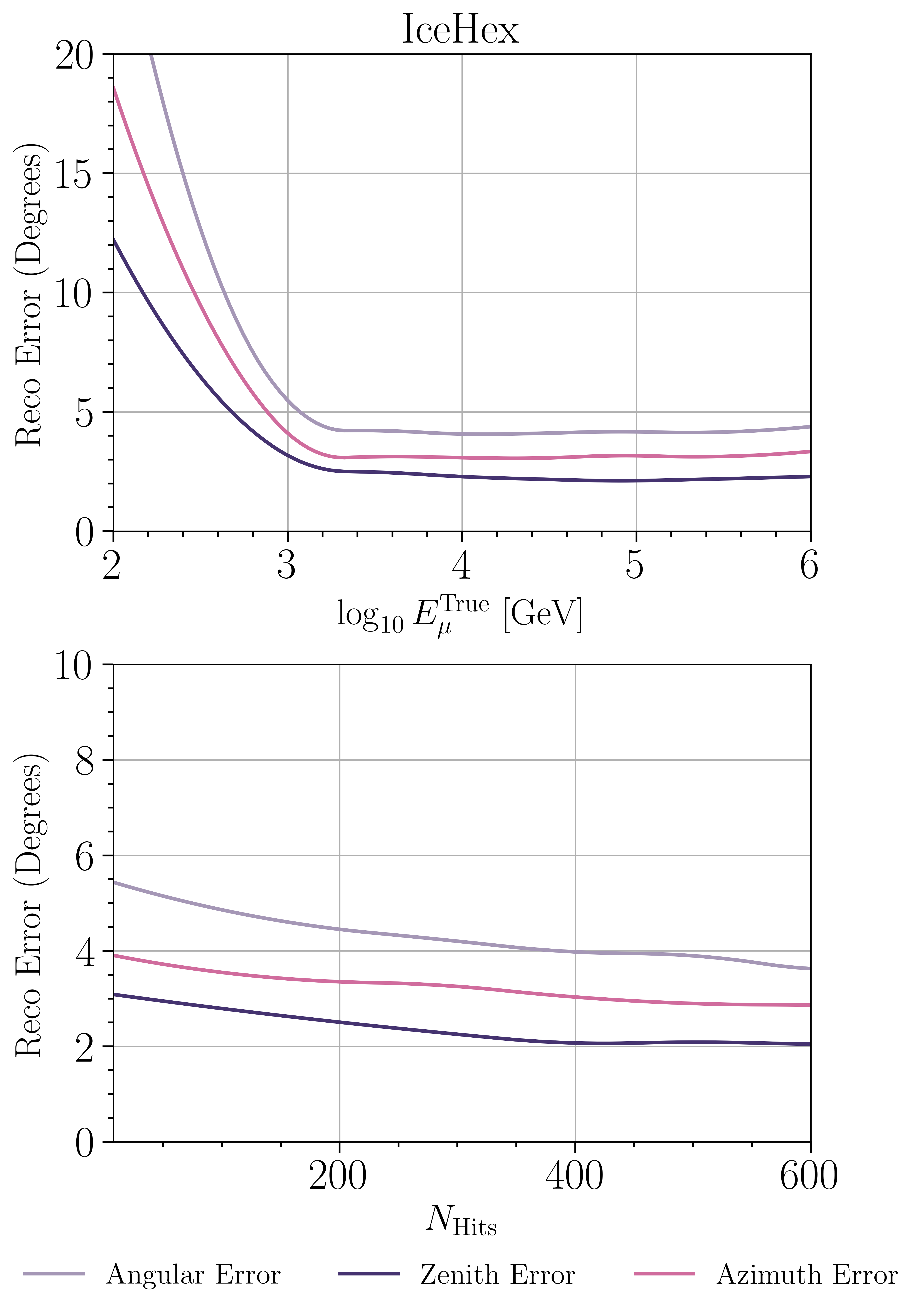}
    \caption{\textbf{Neutrino Angular Reconstruction Resolution on full accuracy network for IceHex a function of the true simulated lepton energy and number of photon hits:} Top panel shows the error as a function of the muon, bottom panel shows that as a function of the number of photon hit deposition in OMs.}
    \label{fig:angular_reco_errors_lepton}
\end{figure}

We first show the accuracy of the network as evaluated on GPU using double-precision floating-point computations, without mapping or imposing the cutoff at $256$ for the entries of the input.
This serves as a benchmark as it is the optimal reference of the algorithm.
In~\Cref{fig:angular_reco_errors}, we show the median error distribution of the angular, zenith, and azimuthal reconstruction against true neutrino energy for both IceHex and WaterHex detectors at trigger level.
We use the same architecture but train on both datasets separately.
For the IceHex detector, we see the angular error reaching as low as $4.0 \degree$; and for the WaterHex detector, we see the angular error reaching as low as $5.6 \degree$.
The performance for IceHex and WaterHex are comparable, where we do observe the detector in ice to have events that are easier to reconstruct, due to the uniform choice of 120~meters as the string spacing across the two mediums, whereas photons in water have a smaller absorption length, resulting in Cherenkov photons deposition in fewer, more clustered OMs.
Additionally, the rising tail on the high energy end is due to event selection at triggering, where higher energies allow for the more poorly positioned track, cutting only the corner of the detector, for example, depositing fewer photons.
On the other hand, as shown in the bottom panel of~\Cref{fig:angular_reco_errors_lepton}, the reconstructed error decreases monotonically with the increment in the number of photons hit depositions in the OMs.
While this network is designed with limitations in architecture to enable further acceleration on TPUs, it is comparable in accuracy performance with other machine learning-based reconstruction algorithms for similar, but not completely identical, detectors~\cite{Yu:2023ehc}~\cite{IceCubeDNN} at the trigger level.
The recent high-speed algorithm using sparse submanifold CNN method~\cite{Yu:2023ehc} at the trigger level is labeled as a dotted gray line in~\Cref{fig:angular_reco_errors}.

\subsection{Post-Quantization Accuracy}

\begin{figure}[h]
    \centering
    \includegraphics[width=0.85\textwidth]{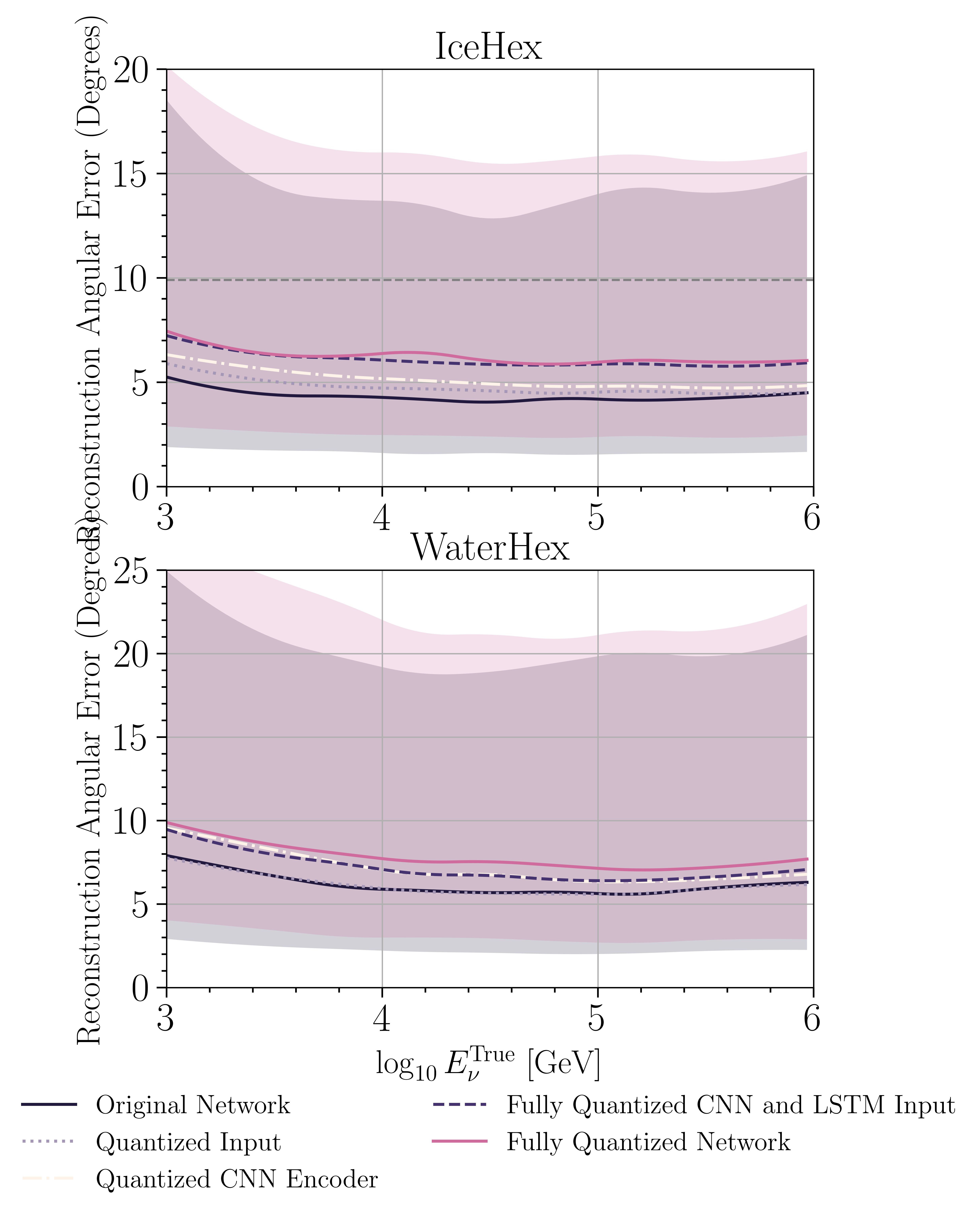}
    \caption{\textbf{Post-Quantization Accuracy of the Network:} Aside from the baseline performance of the original network on the original input, we show the median angular reconstruction error of the network after the quantization of input, input and CNN encoder, and entire network, respectively. Dashed lines show performances at intermediate quantization steps. Colored regions show $20\%$ and $80\%$ containment for the original network and fully quantized network. For the IceHex detector, the dotted gray line shows the IceCube reported median angular error using the LineFit algorithm~\cite{LineFit} at trigger level, the current real-time reconstruction method under the resources restrictions.}
    \label{fig:post_q_acc}
\end{figure}

In this section, we present and discuss the network performance at each stage after the quantization sub-steps described in~\Cref{subsec:quantization} are applied.
We show the energy distribution of angular reconstruction error at various stages, together with the region of top and bottom $20\%$ errors in~\Cref{fig:post_q_acc}.

For the IceHex detector, performance is nearly unaffected by the quantization of inputs, keeping the network in floating point precision.
At this stage, the median error is $4.6\degree$, with the high energy end of the spectrum reaching a $4.4\degree$ median error.
Upon quantization of the CNN encoder, applying a float fallback before inputting into the LSTM cell, which is kept at floating point precision, the median angular error reaches $5.0 \degree$.
Disabling floating point fallback and thereby quantizing the entire CNN encoder as well as the LSTM inputs, upon retraining the LSTM cell, we see the median reconstructed angular error reach $6.0 \degree$. This increment in error is avoidable by bringing in a software that is capable of quantizing the entire network without having to quantize by parts and fine-tune.
Upon quantizing the entire network and evaluating the error after falling back from the integer prediction to their respective original floating point numbers via the reverse quantization mapping, we see the median error reach $6.1\degree$.
This still beats the median angular error of $9.9\degree$ of the currently employed real-time solution employed by IceCube~\cite{LineFit}, which is a CPU-based regression algorithm.
This implies a very well-tuned network that works well with the quantization scheme since this error already accounts for the dead-weight loss that comes with low-power edge computing: the prediction being discrete as opposed to the continuous nature of the simulated ground truth. 

For the WaterHex geometry, we observe a similar behavior across the different stages of quantization.
The median reconstructed angular error being $6.0\degree$, $6.9\degree$, $7.0 \degree$, $7.7\degree$ respectively for the $4$ quantization stages in order.

\begin{figure*}[t]
    \centering
    \includegraphics[width=0.9\linewidth]{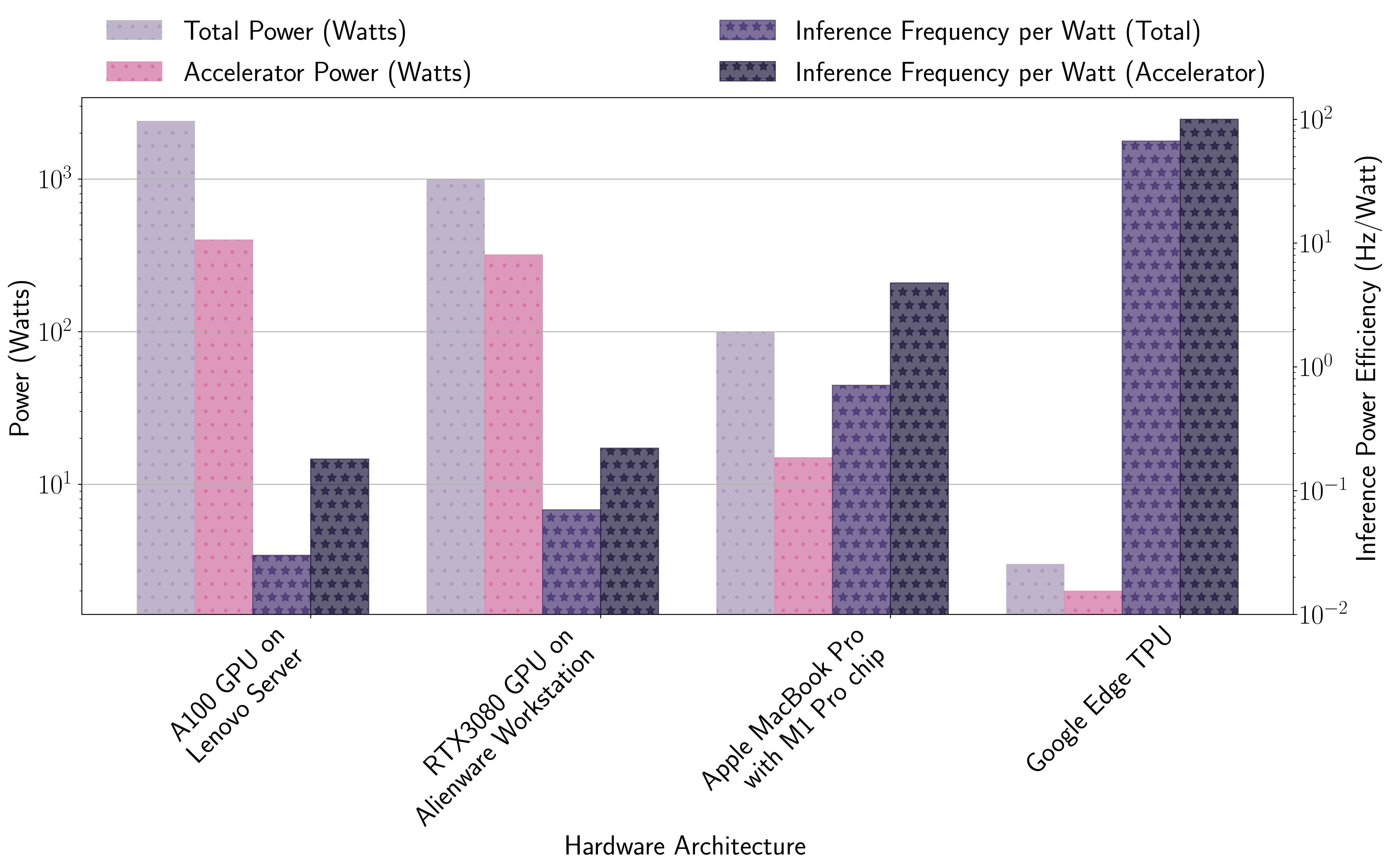}
    \caption{\textbf{Network power efficiency when run on different architectures:} lighter-colored, dotted bars show the total and accelerator power, star-filled, darker-colored bars show the power efficiency in inference frequency per watt.}
    \label{fig:bar}
\end{figure*}
\subsection{Inference Frequency Performance}

With this benchmark network accuracy performance, we test the network on the various hardware architectures and hereby report the inference frequencies.
For the edge-inference performance testing, we compile the fully quantized versions with the \texttt{PyCoral} compiler and deploy them on the Google Edge TPU DevBoard; for the GPU run-time measurements, we run the full precision models on the various GPU architectures.
It is important to note that non-fully quantized versions with TPU-incompatible operations but integer precision computations can be run on GPU architectures with a reduced run-time, but the model developed for this work is specifically designed to be run on the Edge TPU, satisfying its limitations, in order to tackle to power limitation problem.
Therefore, while accelerating machine learning inference on GPUs with quantization can be of interest to future work, in this work we only focus on the comparison between the original performance on GPU with full precision on the one end of the spectrum and fully quantized network performance on TPU.
\Cref{fig:bar} shows how the performance per watt increases as the scale of the computing system decreases for single-event inference.

In a real-time, in-detector environment, for a triggering task, inferences are made on single events: this goes in the opposite direction of the strength of GPUs, which is its speed on large batch size parallel computation.
In~\Cref{fig:bar}, we are considering this scenario of single event inference, which results in the very poor performance of large-scale computing systems towards the left side of the plot, and favors the smaller-sized architectures on the other end of the spectrum.
We observe that for single event inference, the A100 GPU, RTX3080 GPU, and the core GPU on M1 chip demonstrate similar inference speeds at around 13~milliseconds per inference, the Google Edge TPU Dev Board hits 5~milliseconds in inference, reaching 100~Hz/Watt performance while operating at 3~Watts, accounting for the form factor aside from the TPU chip itself.
This enables us to perform machine learning algorithms in real-time inside the detector, where power is limited to, for example for IceCube, 5.7~Watts per optical Module~\cite{IceCube:2016zyt}, while for KM3nET it is 7~Watts per module~\cite{Leonora_2018}.

Our work by no means undermines the capability of large-scale computing systems: if we increase the batch size to 100, then we observe that the time per inference decreases to 0.4 milliseconds for the A100 card, 1.7 milliseconds for the RTX 3080 card, and 3.8 milliseconds for the M1 Pro Chip.
For algorithms with a more efficient data encoding, like that of Sparse-Submanifold CNNs~\cite{Yu:2023ehc}, that enables a very large batch size, A100 cards are capable of performing inference at a rate of 9901 Hz on a batch size of 12288.
Edge TPUs, on the other hand, are incapable of performing inference on a batch, but it is exactly the gains of power efficiency on single events that allows us to enable real-time in-detector machine learning inference using these edge devices.
Thus GPU-designed algorithms and associated hardware are well-suited for off-line event filtering, and particle identification and reconstruction.

\section{Summary and Outlook}\label{sec:Outlook}
In this article, we have shown the accuracy and power efficiency of TPU-tailored deep neural network for water- and ice-Cherenkov neutrino telescope event reconstruction.
Our work serves as a proof of concept for the feasibility of real-time low-power in-situ machine learning tuned for on-going and next generation neutrino experiments.
We have demonstrated the capability of a low-power machine learning algorithm following a specifically designed input pre-processing and quantization procedure, capable of reaching peaking power efficiency while maintaining a competent accuracy, beating non-machine learning algorithms that are currently being deployed in real time reconstructions or trigger systems.
However, our work is far from a realization of the full potential of edge computing.
To begin with, since the outputs are quantized, angular reconstruction, a continuous value prediction problem in nature, is not an optimal problem for such an algorithm to tackle.
Alternatively, classification problems are better suited for deployment on edge computing, which is a future direction left for work.
Additionally, the field of edge computing is under very fast development on both the software and the hardware fronts, some recent developemnts include the recent release of the \texttt{PyTorch} \texttt{ExecuTorch} that enables a new end-to-end solution for edge computing, supporting more edge devices~\cite{executorch}.

Looking forward, for next generation detectors, a new variety of situations will appear that calls out for the need of a low-power real-time machine-learning based data handling, including but not limited to the following scenarios:
\begin{itemize}
    \item \textit{Multiple-PMT Optical Modules}: Among designs of next generation optical modules, many have incorporated multiple PMT's into single modules~\cite{anderson2021design}.
    This opens up the space for an algorithm that processes the multiple waveforms of this single module as a time series data.
    In this scenario, the incorporation of an edge computing device would enable us to deploy machine learning based sophisticated algorithms assisting this processing of local data.
    \item \textit{Real Time Triggering}: Next generation neutrino telescopes will typically be much larger in geometric and effective areas, looking at IceCube-Gen2, for example~\cite{IceCube-Gen2:2020qha}.
    This implies a larger amount of data transmission and storage requirement if we keep using the same simple cutoff-like triggers.
    Inclusion of real-time machine learning capability will allow us to instead develop machine learning based triggering, which will not only help us in detecting rear event in real time, but also assist us in data selection and therefore alleviate the stressed data transmission and storage system.
    \item \textit{Other power-limited facilities}: Aside from water-/ice- Cherenkov neutrino telescopes, there are other experiments in similarly, extremely, power-limited environments, such as satellite detectors~\cite{Litebird}.
    Enabling machine learning in such environments will allow for a real time data handling system in these scenarios as well.
    \item \textit{Other edge computing devices}: There are many more edge computing alternatives that are low-power and efficient, and they usually face the same set of restrictions.
    Using similar quantization ideas developed in this work, we can explore more variations of micro-architectures, evaluating the pros and cons of each before incorporating them into the next generation intelligent detector hardware.
\end{itemize}

As such, through this first demonstration of machine learning effort on TPUs, we would like to motivate similar further exploration into this direction of low-power computing alternatives.
There are definitely various other improvements and applications waiting out there to be explored along this newly opened up gateway.

\section*{Code Availability}
The code used to train the network and produce the plots in this work can be found in \href{https://github.com/MiaochenJin/RecoOnEdge}{GitHub Repository}.

\begin{acknowledgments}
We thank Simone Francescato for painstaking comments and suggestions. MJ would also like to thank Nicholas Kamp, Felix Yu, Lihao Yan, Yidi Qi and Jinzheng Li for useful discussions. CAA is supported by the Faculty of Arts and Sciences of Harvard University and the National Science Foundation (NSF).
Through part of this work, they were also supported by the Alfred P. Sloan Foundation.
MJ was supported by the Harvard Physics Department Purcell Fellowship for part of this work.
The NSF partially supported this work under Cooperative Agreement PHY-2019786 (The NSF AI Institute for Artificial Intelligence and Fundamental Interactions, http://iaifi.org/)
Finally, CAA and MJ are supported by NSF CAREER Award PHY-2239795.
\end{acknowledgments}


\nocite{*}
\newpage
\bibliography{RecoTPU}
\clearpage


\newpage

\onecolumngrid
\appendix

\ifx \standalonesupplemental\undefined
\setcounter{page}{1}
\setcounter{figure}{0}
\setcounter{table}{0}
\setcounter{equation}{0}
\fi

\renewcommand{\thepage}{Supplemental Methods and Tables -- S\arabic{page}}
\renewcommand{\figurename}{SUPPL. FIG.}
\renewcommand{\tablename}{SUPPL. TABLE}

\renewcommand{\theequation}{A\arabic{equation}}

\section{Network Architecture}\label{appx:NetworkArchi}

Here we show in detail the network architecture developed in this work.
Firstly, in~\Cref{table:ResNetBlock} we show a cellular basic component of the CNN encoder: the residual convolution block. This block is used extensively in the CNN encoder, whose architecture and hyperparameters are shown in~\Cref{table:CNNEncoder}.
We finally show the entire network architecture, consisting of tensor reshaping layers, the CNN encoder, and the LSTM block together with a final fully connected layer in~\Cref{table:Network}.

\begin{table*}[b]
\begin{tabular}{l|c}
\makecell{Layer Name} & \makecell{Layer Specs} \\
\hline \hline

\makecell{$\texttt{Input}$}	& $X_{\mathrm{in}}$ \\
\hline
\makecell{$\texttt{Conv2D}_1$}	& Kernel Size $(k_x, k_y)$; Stride $s_1$; Channels $c$; Padding\\
\hline
\makecell{$\texttt{Conv2D}_2$}	& Kernel Size $(k_x, k_y)$; Stride $s_2$; Channles $c$; Padding\\
\hline
\makecell{$\texttt{Addition}$}	& \texttt{Add}$(X_{\mathrm{in}} + \texttt{Conv2D}_1(\texttt{Conv2D}_2(X_{\mathrm{in}})))$\\
\end{tabular}
\caption{\textbf{Residual Block:} architecture of a single cellular residual block with given kernel size $(k_x, k_y)$, stride $(s_1, s_2)$, and some padding strategy, used extensively in the CNN encoder of the network. $s_1 \neq s_2$ implies downsampling.}
\label{table:ResNetBlock}
\end{table*}

\begin{table*}[b]
\begin{tabular}{l|c}
\makecell{Layer Name} & \makecell{Layer Specs} \\
\hline \hline

\makecell{$\texttt{Input}$}	& $X_{\mathrm{in}}$\\
\hline
\makecell{$\texttt{Conv2D}_1$}	& Kernel Size $(3, 3)$; Stride $2$; Channels: $32$; Padding: $\texttt{"same"}$\\
\hline
\makecell{$\texttt{ResBlock}_1$}	& $k_x = k_y = 3$; $s_1 = s_2 = 1$; $c = 32$; Padding  $=\texttt{"same"}$\\
\hline
\makecell{$\texttt{ResBlock}_2$}	& $k_x = k_y = 3$; $s_1 = 2, s_2 = 1$; $c = 32$; Padding $=\texttt{"same"}$\\
\hline
\makecell{$\texttt{ResBlock}_3$}	& $k_x = k_y = 3$; $s_1 = s_2 = 1$; $c = 64$; Padding $=\texttt{"same"}$\\
\hline
\makecell{$\texttt{ResBlock}_4$}	& $k_x = k_y = 3$; $s_1 = 2, s_2 = 1$; $c = 64$; Padding $=\texttt{"same"}$\\
\hline
\makecell{$\texttt{ResBlock}_5$}	& $k_x = k_y = 3$; $s_1 = s_2 = 1$; $c = 128$; Padding $=\texttt{"same"}$\\
\end{tabular}
\caption{\textbf{CNN Encoder:} architecture of the CNN encoding network that encodes the input at any time step $t$: $A_t^{(X,Y,Z)}$ into a 1 dimensional vector in the hidden latent space, which in turn gets processed by the LSTM Block.}
\label{table:CNNEncoder}
\end{table*}

\begin{table*}[b]
\begin{tabular}{l|c}
\makecell{Layer Name} & \makecell{Layer Specs} \\
\hline \hline

\makecell{$\texttt{Input}$}	& $\{A_T^{(X, Y, Z)}\}_B$\\
\hline
\makecell{$\texttt{Reshape}_1$}	& $\texttt{Reshape}(\{A_T^{(X, Y, Z)}\}_B, [B \times T, X, Y, Z])$\\
\hline
\makecell{$\texttt{CNN Encoder}$}	& See~\Cref{table:ResNetBlock}\\
\hline
\makecell{$\texttt{Reshape}_2$}	& $\texttt{Reshape}(\texttt{CNN Encoder}(\texttt{Reshape}(\{A_T^{(X, Y, Z)}\}_B, [B \times T, X, Y, Z])), [B, T, 128])$\\
\hline

\makecell{$\texttt{LSTM}$}	& LSTM Dimension: $128$; $n_{\mathrm{hidden}}$: 128\\
\hline
\makecell{$\texttt{Dense}$}	& To Output\\

\end{tabular}
\caption{\textbf{Network Architecture:} architecture of the entire network, with given input sizes depending on the data pre-procerssing and detector geometry parameters $T, X, Y, Z$.}
\label{table:Network}
\end{table*}
\end{document}